\documentclass{bioinfo}

\usepackage{graphicx}

\copyrightyear{XXXX}
\pubyear{XXXX}

\usepackage{natbib}
\usepackage{url}
\usepackage{alltt}
\usepackage{subcaption}
\usepackage{todonotes}
\usepackage{verbatimbox}
\usepackage{soul}
\usepackage{float}

\begin{document}
\firstpage{1}

\title[Ococo: an online variant and consensus caller]{Ococo: an online variant and consensus caller}

\author[K. Břinda\ \textit{et~al.}]{
	Karel Břinda\,$^{1,2}$\footnote{to whom correspondence should be addressed},
	Valentina Boeva\,$^{3,4}$, and
	Gregory Kucherov\,$^{5,6}$
}

\address{
	$^{1}$Center for Communicable Disease Dynamics, Department of
	Epidemiology, Harvard TH Chan School of Public Health, Boston MA;
	$^{2}$Department of Biomedical Informatics,
	Harvard Medical School, Boston MA;
	$^{3}$Institut Curie -- Centre de Recherche, PSL Research University,
	Mines Paris Tech, INSERM U900, Paris, France;
	$^{4}$Institut Cochin, INSERM U1016, CNRS UMR 8104, Université
	Paris Descartes UMR-S1016, Paris, France;
	$^{5}$LIGM/CNRS, Université Paris-Est, Marne-la-Vallée, France;
	$^{6}$Skolkovo Institute of Science and Technology (SkolTech), Moscow Region, Russia.
}

\history{Received on XXXXX; revised on XXXXX; accepted on XXXXX}
\editor{Associate Editor: XXXXXXX}
\maketitle

\begin{abstract}

\section{Motivation:}
Identifying genomic variants is an essential step for connecting genotype and phenotype. The usual approach consists of statistical inference of variants from alignments of sequencing reads. State-of-the-art variant callers can resolve a wide range of different variant types with high accuracy. However, they require that all read alignments be available from the beginning of variant calling and be sorted by coordinates. Sorting is computationally expensive, both memory- and speed-wise, and the resulting pipelines suffer from storing and retrieving large alignments files from external memory. Therefore, there is interest in developing methods for resource-efficient variant calling.

\section{Results:}
We present \textsc{Ococo}, the first program capable of inferring variants in a real-time, as read alignments are fed in. \textsc{Ococo} inputs unsorted alignments from a stream and infers single-nucleotide variants, together with a genomic consensus, using statistics stored in compact several-bit counters. \textsc{Ococo} provides a fast and memory-efficient alternative to the usual variant calling. It is particularly advantageous when reads are sequenced or mapped progressively, or when available computational resources are at a premium.

\section{Availability:} \url{http://github.com/karel-brinda/ococo}

\section{Contact:}
\url{kbrinda@hsph.harvard.edu}
\end{abstract}

\section{Introduction}

Identifying genomic variants is an essential step for connecting genotype and phenotype. The goal of \emph{variant calling} is to identify genomic variants present in the sequenced individual or a population. Most commonly, variant calling proceeds by read mapping and then sliding a small window throughout the genome, collecting statistics for all reads aligned within the window and calculating the likelihood of variants observed in these alignments. We term this approach \emph{offline variant calling} as it requires that all read alignments are available from the beginning. Offline calling is implemented in all major variant callers (see, e.g., \citep{Bao2014}).

However, offline variant calling is highly time- and space-demanding. First, all alignments must be available and get sorted by coordinates prior to variant calling; this involves storing and retrieving large alignment files from external memory. Second, variant callers usually apply computationally expensive steps, such as realignments, even for regions where this is not necessary. The resulting performance can be particularly limiting on portable devices, personal computers, or in a cloud environment with restricted resources.

Here, we introduce the concept of \emph{online variant calling}, where variants are inferred in real time, as read alignments are fed in. We implement this approach in a program called \textsc{Ococo}, the first online variant caller. \textsc{Ococo} inputs unsorted alignments from an unsorted SAM/BAM stream \citep{Li2009} and infers single-nucleotide variants, together with a genomic consensus, using statistics stored in compact several-bit counters. \textsc{Ococo} provides a fast and memory-efficient alternative to the usual variant calling, which is particularly advantageous when reads are sequenced or mapped progressively, or when available computational resources are at a premium.

\section{Methods}

\begin{methods}

\textbf{Overview.} \textsc{Ococo} calls variants and consensus directly from an unsorted SAM/BAM file, possibly provided in a stream. To do that, \textsc{Ococo} stores and maintains variant statistics for all genomic positions about previous alignments as well as a consensus sequence, which represents the current internal reference. The consensus can be initialized from a user-provided sequence, typically the same as used for read mapping. Whenever a new alignment is loaded, \textsc{Ococo} updates the statistics and assesses whether they are still concordant with the consensus. If not, the consensus is corrected and the corresponding substitution reported as a novel variant.

\textbf{Compact representation of variant statistics.} In the online approach reads can potentially map to any location. This is a fundamental difference from the offline calling, where reads are sorted and statistical inference uses a small sliding window, collecting information about locally overlapping alignments. Therefore, the main challenge of online calling is to design variant statistics for the whole genome that fit into main memory and at the same time be sufficiently informative for inferring variants. We propose using small, several-bit nucleotide counters and complementing them with fast bit operations.

The \textsc{Ococo} statistics consist of four integer counters per position, one per each nucleotide (\textbf{Figure 1}). Every counter represents the number of nucleotides aligned to that position; however only the most significant bits are stored. Whenever a new alignment is loaded, the corresponding counters are incremented (\textbf{Figure 1a}). If a counter is already saturated and yet is to be incremented, then all counters at the position are first bit-shifted, losing their rightmost (least significant) bit (\textbf{Figure 1c}). This mechanism makes it possible to compute nucleotide frequencies in a limited space and filter out randomly distributed sequencing errors. \textsc{Ococo} supports three counter configurations: 16, 32, and 64 bits per position corresponding to 4 bits for the consensus base and 3, 7, and 15 bits for each nucleotide counter, respectively.

\textbf{Variant and consensus calling strategy.} For the sake of speed, we propose considering individual genomic positions independently and keeping the consensus synchronized with the dominating base if such exists (\textbf{Figure 1b}). Formally defined, let \(C_{n}\) denote the value of the counter for a nucleotide \(n \in \{ A,C,G,T\}\) at some fixed position, then the consensus is updated to \(n\) if \({2C}_{n} > C_{A} + C_{C} + C_{G} + C_{T}\). This represents a simple instantiation of maximum likelihood estimation for haploid genomes and single-nucleotide variants without considering base qualities. Variants and the resulting consensus are reported in the VCF \citep{Danecek2011} and FASTA format, respectively.

\textbf{Working modes.} \textsc{Ococo} supports two modes of online calling: the \emph{real-time} and \emph{batch modes}. Whereas the \emph{real-time mode} updates consensus and reports variants immediately after processing each read, the \emph{batch mode} postpones reporting updates until all reads from the current batch have been processed.

\textbf{Implementation.} \textsc{Ococo} is implemented in C++ and released under the MIT license. The software package is available from \url{http://github.com/karel-brinda/ococo}, BioConda \citep{Gruning2018}, and Zenodo \citep{Brinda2017}.

\end{methods}

\begin{figure}
	\includegraphics[width=\linewidth]{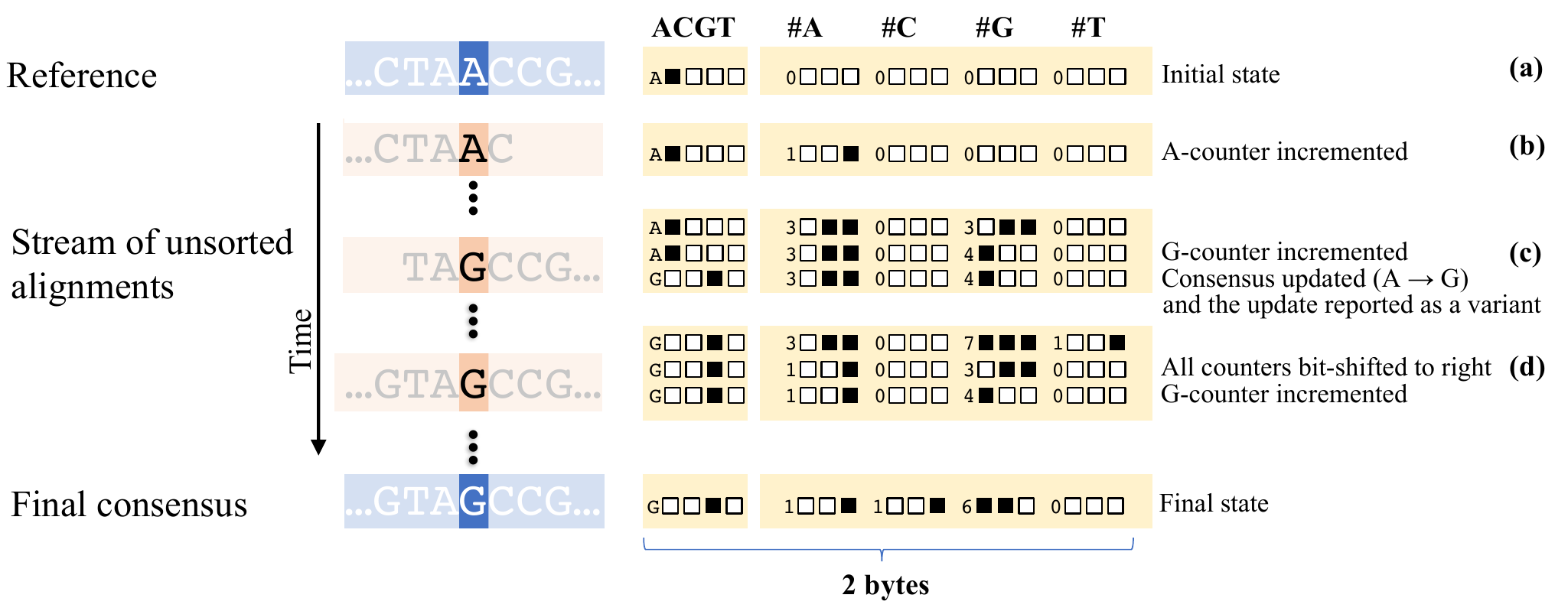}
	\caption*{\textbf{Figure 1. Internal statistics of an online variant caller.} Example of update of \textsc{Ococo} counters for a single position of the genome. The first 4 bits carrying the nucleotide consensus are followed by 4 nucleotide counters, each of them 3 bits long in this case. Vertical axis corresponds to time. The figure shows how the counters and consensus are updated based on the received alignments. \textbf{a)} At the beginning, the consensus base is initialized to the reference base (A). All counters are set to 0. \textbf{b)} The A counter is incremented; the statistics stay concordant with the consensus. \textbf{c)} The G counter is incremented, which triggers a consensus update and reporting a new variant (A$\rightarrow$G). \textbf{d)} The G counter is to be incremented, but it is already saturated. Therefore, all counters of the position must be bit-shifted first.}
\end{figure}

\section{Results}

We performed two experiments demonstrating the accuracy and resource-efficiency of \textsc{Ococo} (\textbf{Supplementary Figure 1}). Using \textsc{RnfTools} (v0.3.1.3) \citep{Brinda2016b} with \textsc{DwgSim} (v0.1.11), we simulated reads of length 100bp from the \emph{Chlamydia trachomatis} genome (NC\_021897.1, 1.046Mbp), with 20x coverage and 1\% sequencing error rate. The genome was \emph{in silico} mutated at the 2\% level (with 15\% of mutations being indels and with the 30\% indel extension probability). The reads were then mapped back to the original reference using \textsc{Bwa-Mem} (v0.7.17) \citep{Li2013}.

First, we evaluated how online variant calling progressed in time (\textbf{Supplementary Figure 1a}). We used the obtained alignments to call SNPs in the \textsc{Ococo} (v0.1.2.6) streaming mode with 7-bit counters. After processing the first 50 thousand reads (corresponding to approximately 5x coverage), \textsc{Ococo} neared a plateau of the edit distance between the simulated and inferred genomes. At this point, \textsc{Ococo} had correctly identified 16,239 out of 21,914 single-edit variants (17,613 SNPs; indels of total length 4,301) inserted into the reference genome, i.e., 92,2\% of SNPs were identified. After all the 209,742 reads have been processed, the proportion of correctly identified SNPs increased to 97,3\%.

Second, we compared the speed of the \textsc{Ococo} batch mode with a common variant-calling pipeline consisting of \textsc{SamTools} (v1.9) \citep{Li2009} and \textsc{VarScan} (v2.4.3) \citep{Koboldt2009} (\textbf{Supplementary Figure 1b}). For this pipeline, we measured the time required for sorting the alignments, computing an alignment pileup, and for the subsequent SNP calling. Times were measured in seconds, as the mean over 3 runs, using \textsc{SnakeMake} (v5.3.0) \citep{Koster2012} on an iMac 4.2 GHz Intel Core i7 with an SSD disk and 40 GB RAM. We observed that \textsc{Ococo} provided 66x speedup for calling variants compared to \textsc{SamTools} and \textsc{VarScan}.

To assess scaling to larger genomes, we ran the same experiments on human chromosome 17 (HG18, 78,775Mbp) (\textbf{Supplementary Figure 2}). The obtained results were qualitatively similar: 91,3\% and 96,1\% of SNP were identified with 5x and 20x of reads, respectively, and \textsc{Ococo} provided 55x speedup. The decrease in speed-up can be explained by more CPU cache misses due to the higher number of counters for longer genomes.

\section{Discussion}

\textsc{Ococo} brings several advantages over offline variant callers. First of all, it enables determining variants and consensus sequences directly from the output of a read mapper, avoiding heavy I/O operations. In many applications, \textsc{Ococo} produces a sufficiently good-quality consensus supplemented by information about nucleotide frequencies at each position. This is especially relevant for bacterial genomics, where many statistical methods consider single-nucleotide variants only. Finally, online consensus calling is also an essential component of the \emph{dynamic mapping approach} that we developed in a separate study \citep{Brinda2016a}.

A key ingredient of \textsc{Ococo} is the memory-efficient alignment statistics featuring small counters. We observe that they can be seen as an instance of approximate counting \citep{Morris1978}, but with a modified formulation: whereas traditional algorithms estimate \emph{counts}, variant calling requires estimating \emph{ratios} of the counts of nucleotides. Here, we provide an algorithmic description of the counter mechanics, although statistical properties of the resulting estimators are yet to be studied.

A major limitation of \textsc{Ococo} is its restriction to single-nucleotide variants. Indels and more complicated variants present three challenges for future research. First, it is necessary to design an appropriate statistics, ideally counter-based, for each new variant type. Second, the statistics must be complemented with rules for triggering an update, which should be fast to evaluate. Finally, non-substitution updates entail changes in genomic and counter coordinates, which calls for a more sophisticated addressing and allocation system than the one implemented in \textsc{Ococo}.

\section{Acknowledgements}

The work has been supported by ABS4NGS {[}ANR-11-BINF-0001{]}, Labex Bézout, and The Bill and Melinda Gates Foundation {[}GCGH GCE, OPP1151010{]}. VB was supported by the ATIP-Avenir Program, the ARC Foundation {[}RAC16002KSA-R15093KS{]}, and the ``Who Am I?'' laboratory of excellence {[}ANR-11-LABX-0071{]} funded by the French Government {[}ANR-11-IDEX-0005-02{]}. The authors would like to thank Victoria E Jones, Cristina M Herren, and Siân V Owen for critical reading and helpful suggestions.

\paragraph{Conflict of Interest\textcolon}
None declared.

\bibliography{bibliography}

\clearpage{}

\begin{figure*}[h!]
	\begin{tabular}{p{0.5\linewidth}p{0.5\linewidth}}
		\begin{center}
			\textbf{a)}
			\newline
			\includegraphics[width=0.85\linewidth]{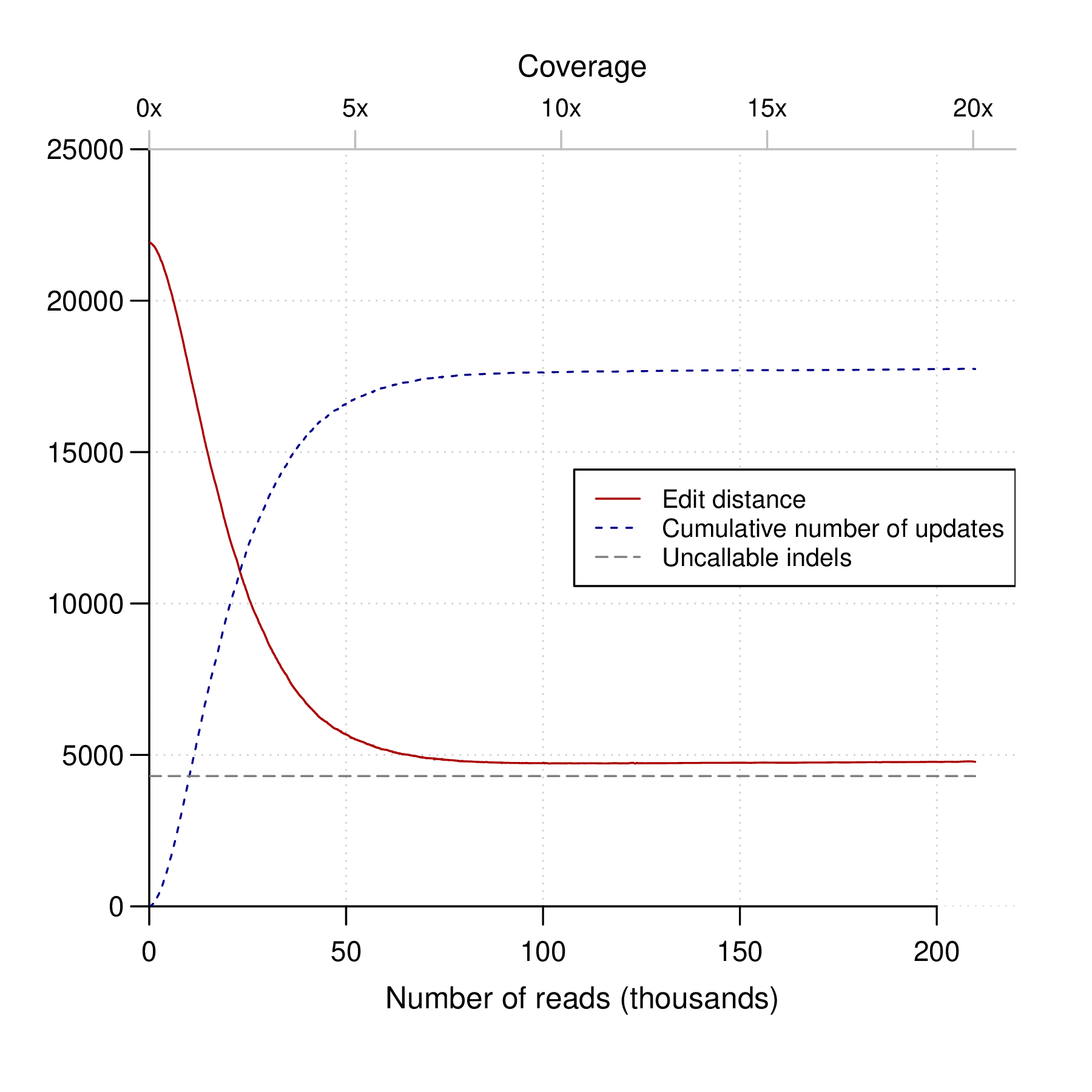}
		\end{center}
		&
		\begin{center}
			\textbf{b)}
			\newline
			\includegraphics[width=0.85\linewidth]{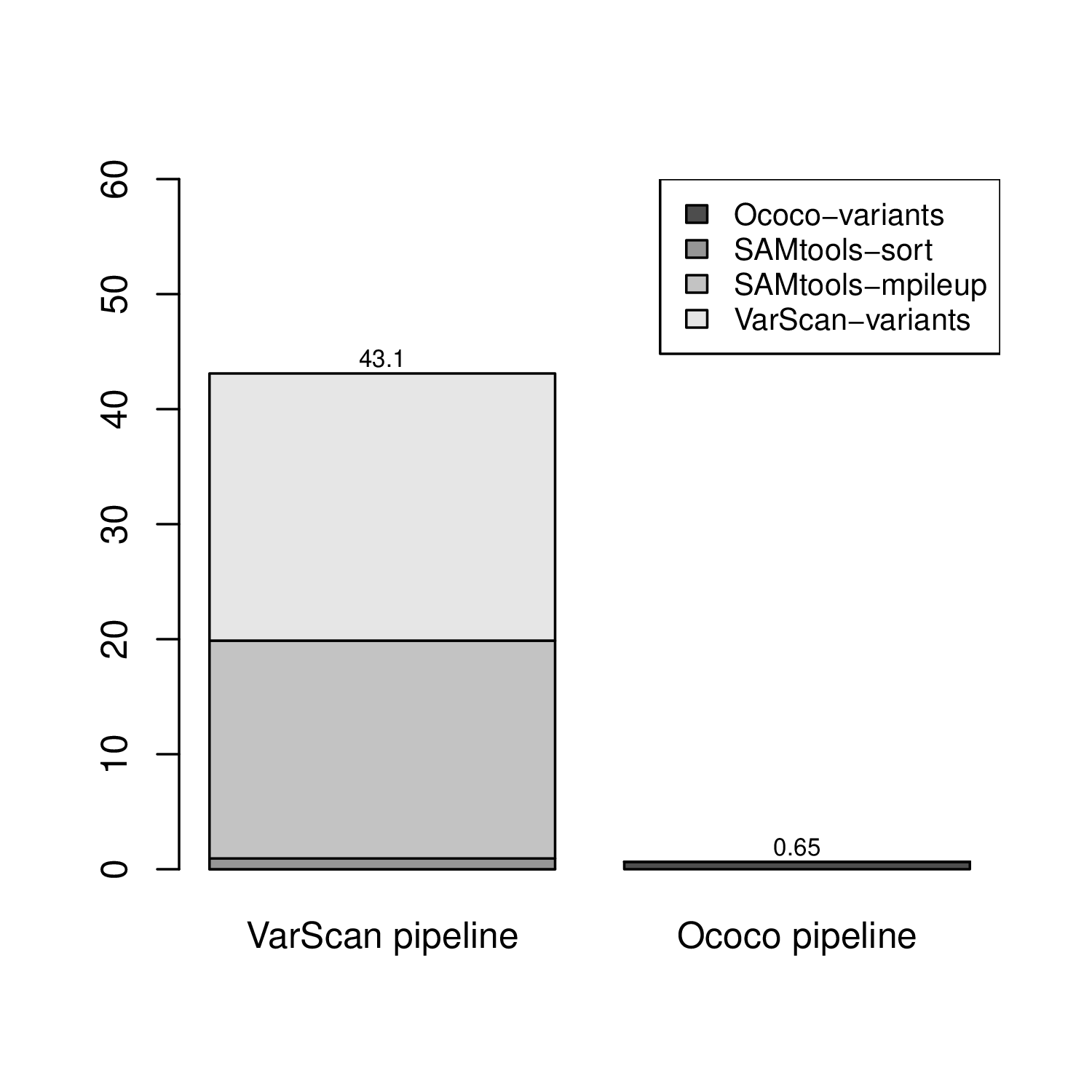}
		\end{center}
	\end{tabular}
	\caption*{\textbf{Supplementary Figure 1. Evaluation of \textsc{Ococo} with Chlamydia Trachomatis (1.046Mbp)}. \textbf{a) Online variant calling as a function of time.} The blue curve shows the cumulative number of updates of the consensus as a function of the number of processed alignments (or the actual coverage). The red curve shows the edit distance from the simulated sequenced genome. \textbf{b) Speed comparison.} Comparison of time to completion of variant calling using \textsc{Ococo} and a pipeline based on \textsc{VarScan}.}
\end{figure*}

\begin{figure*}[h!]
	\begin{tabular}{p{0.5\linewidth}p{0.5\linewidth}}
		\begin{center}
			\textbf{a)}
			\newline
			\includegraphics[width=0.85\linewidth]{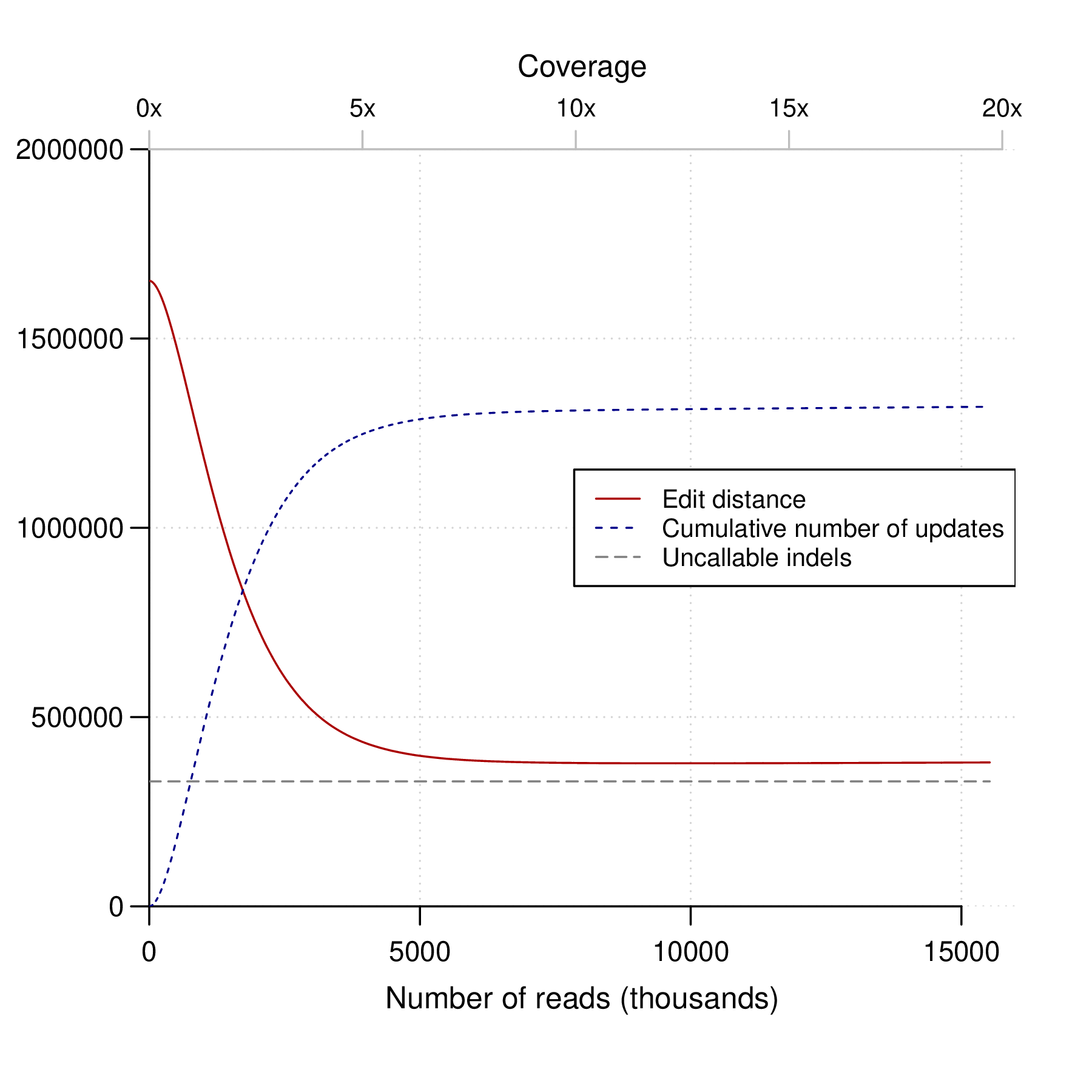}
		\end{center}
		&
		\begin{center}
			\textbf{b)}
			\newline
			\includegraphics[width=0.85\linewidth]{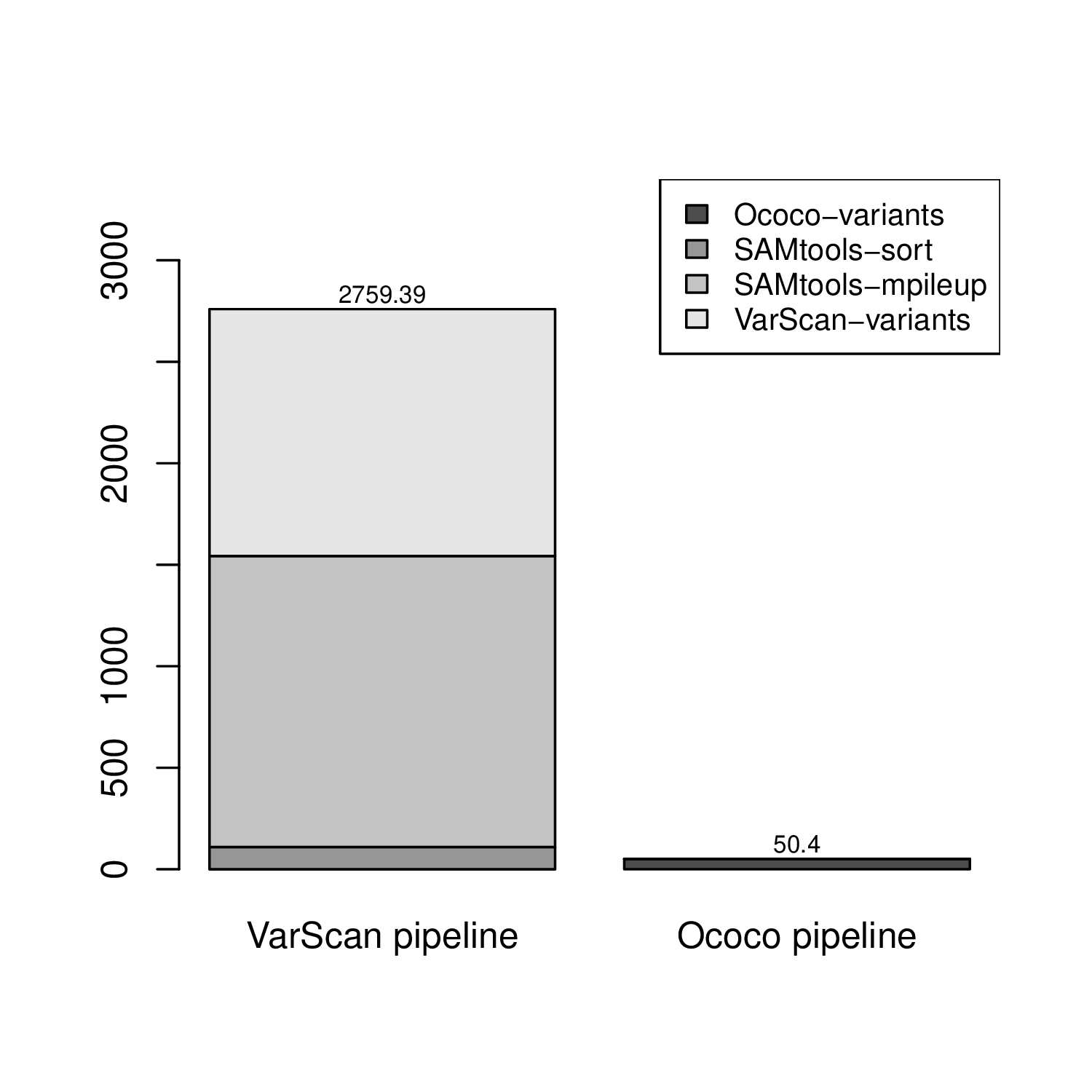}
		\end{center}
	\end{tabular}
	\caption*{\textbf{Supplementary Figure 2. Evaluation of \textsc{Ococo} with human chromosome 17 (78,775Mbp).} The figure is of the same format as \textbf{Supplementary Figure 1.}}
\end{figure*}

\end{document}